\documentclass[aps,prl,twocolumn,groupedaddress,showpacs,showkeys]{revtex4}
\usepackage{graphicx}% Include figure files
\usepackage{dcolumn}% Align table columns on decimal point
\usepackage{bm}% bold math

\begin{document}

%%%%%%%%%%%%%%%%%%%%%%%%%%%%%%%%%%%%%%%%%%%%%%%%%%%%%%%%%%%%
% title/author/abstract

\title{Stripe Charge Ordering in Triangular-Lattice Systems}

\author{Hiroaki Onishi and Takashi Hotta}

\affiliation{%
Advanced Science Research Center,
Japan Atomic Energy Research Institute,
Tokai, Ibaraki 319-1195, Japan}

%\date{\today}
\date{August 18, 2005}

\begin{abstract}
We investigate the ground-state properties of
a $t_{\rm 2g}$-orbital Hubbard model on a triangular lattice
at electron density 5.5 by using numerical techniques.
There appear several types of paramagnetic phases,
but we observe in common that
one or two orbitals among three orbitals become relevant
due to the effect of orbital arrangement.
It is found that charge stripes stabilized by
the nearest-neighbor Coulomb interaction consist of
antiferromagnetic/ferro-orbital chains for small Hund's coupling,
while there occurs stripe charge ordering with
ferromagnetic/antiferro-orbital chains for large Hund's coupling.
\end{abstract}

\keywords{%
$t_{\rm 2g}$-orbital Hubbard model,
triangular lattice,
charge stripe}

\pacs{%
71.30.+h, 71.10.Fd, 75.30.Kz}

% 71.30.+h
% Metai-insulator transitions and other electronic transitions
% 71.10.Fd
% Lattice fermion models (Hubbard model, etx)
% 75.30.Kz
% Magnetic phase boundaries
% (including magnetic transition, metamagnetism, etx)

\maketitle

%%%%%%%%%%%%%%%%%%%%%%%%%%%%%%%%%%%%%%%%%%%%%%%%%%%%%%%%%%%%
% main text

In recent years, a wide variety of exotic phenomena
in layered cobalt oxides have attracted much attention
in the research field of condensed-matter physics.
In particular, the discovery of superconductivity in
water-intercalated cobaltite
Na$_{0.35}$CoO$_{2}$$\cdot$1.3H$_{2}$O \cite{Takada2003}
has triggered current high activity
on the study of superconductivity
in frustrated triangular-lattice systems.
On the other hand,
magnetic transitions have also been observed in
Na$_{x}$CoO$_{2}$ \cite{Motohashi2003,Sugiyama2004}.
In such materials, the competition between various types of
ordered states should arise due to geometrical frustration.
In addition, it has been stressed that
the $t_{\rm 2g}$-orbital degree of freedom plays a crucial role
in causing the rich phase diagram in the cobaltites.
In fact, multiorbital models have been investigated
to understand the mechanism of exotic superconductivity and
novel magnetism in triangular-lattice cobalt
oxides \cite{Mochizuki2005,Koshibae2003,Indergand2005,Mizokawa2004}.
%oxides [4-7].
To gain a deep insight into the behavior of
such complex cobalt-based systems,
it is important to clarify the characteristics of
spin-charge-orbital states
taking account of strong electron correlations

In this paper, we study the ground-state properties of
the $t_{\rm 2g}$-orbital Hubbard model on a triangular lattice
at electron density 5.5.
The Hamiltonian is given by
\begin{eqnarray}
 H &=&
 \sum_{{\bf i},{\bf a},\gamma,\gamma',\sigma}
 t_{\gamma\gamma'}^{\bf a}
 d_{{\bf i}\gamma\sigma}^{\dag} d_{{\bf i}+{\bf a}\gamma'\sigma}
 \nonumber\\
 &&
 -(\Delta/3) \sum_{\bf i}
 (2\rho_{{\bf i}xy}-\rho_{{\bf i}yz}-\rho_{{\bf i}zx})
 +V \sum_{\langle {\bf i},{\bf j} \rangle}
 \rho_{\bf i} \rho_{\bf j}
 \nonumber\\
 &&
 +U \sum_{{\bf i},\gamma}
 \rho_{{\bf i}\gamma\uparrow} \rho_{{\bf i}\gamma\downarrow}
 +(U'/2) \sum_{{\bf i},\sigma,\sigma',\gamma\neq\gamma'} 
 \rho_{{\bf i}\gamma\sigma} \rho_{{\bf i}\gamma'\sigma'}
 \nonumber\\
 &&
 +(J/2) \sum_{{\bf i},\sigma,\sigma',\gamma\neq\gamma'} 
 d_{{\bf i}\gamma\sigma}^{\dag} d_{{\bf i}\gamma'\sigma'}^{\dag}
 d_{{\bf i}\gamma\sigma'} d_{{\bf i}\gamma'\sigma}
 \nonumber\\
 &&
 +(J'/2) \sum_{{\bf i},\sigma\neq\sigma',\gamma\neq\gamma'} 
 d_{{\bf i}\gamma\sigma}^{\dag} d_{{\bf i}\gamma\sigma'}^{\dag}
 d_{{\bf i}\gamma'\sigma'} d_{{\bf i}\gamma'\sigma},
\end{eqnarray}
where $d_{{\bf i}\gamma\sigma}$ is the annihilation operator
for an electron
with spin $\sigma$
in orbital $\gamma$ (=$xy,yz,zx$)
at site {\bf i},
$\rho_{{\bf i}\gamma\sigma}$=
$d_{{\bf i}\gamma\sigma}^{\dag}d_{{\bf i}\gamma\sigma}$,
$\rho_{{\bf i}\gamma}$=$\sum_{\sigma}\rho_{{\bf i}\gamma\sigma}$,
and $\rho_{\bf i}$=$\sum_{\gamma,\sigma}\rho_{{\bf i}\gamma\sigma}$.
Note that the triangular lattice is arranged in the ($x,y$) plane
(see Fig.~1).
The hopping amplitudes are given by
$t_{xy,xy}^{\bf x}=(dd\pi)$,
$t_{yz,yz}^{\bf x}=t_{yz,zx}^{\bf x}=t_{zx,yz}^{\bf x}=0$,
$t_{zx,zx}^{\bf x}=(dd\pi)$
for the ${\bf x}$ direction,
$t_{xy,xy}^{\bf u}=(9/16)(dd\sigma)+(1/4)(dd\pi)$,
$t_{yz,yz}^{\bf u}=(3/4)(dd\pi)$,
$t_{zx,zx}^{\bf u}=(1/4)(dd\pi)$,
$t_{yz,zx}^{\bf u}=t_{zx,yz}^{\bf u}=(\sqrt{3}/4)(dd\pi)$
for the ${\bf u}$ direction,
and
$t_{\gamma,\gamma'}^{\bf v}=t_{\gamma,\gamma'}^{\bf u}$
except for
$t_{yz,zx}^{\bf v}=t_{zx,yz}^{\bf v}=(-\sqrt{3}/4)(dd\pi)$
for the ${\bf v}$ direction.
Note that there is no hopping
between the $a_{\rm 1g}$ ($xy$) and $e_{\rm g}'$ ($yz$ and $zx$) orbitals
due to their different symmetry.
Hereafter, we take $(dd\pi)$=1 as the energy unit.
$\Delta$ is the level splitting
between the $a_{\rm 1g}$ and $e_{\rm g}'$ orbitals,
and $V$ is the nearest-neighbor Coulomb interaction.
In the local interactions, $U$, $U'$, $J$, and $J'$ are
intraorbital Coulomb, interorbital Coulomb, exchange,
and pair hopping interactions, respectively.
Note the well-known relations $J$=$J'$ and $U$=$U'$+$J$+$J'$.
We analyze the model
for eight-site ($N$=8) systems with periodic boundary conditions
by the Lanczos diagonalization.
In this paper, we set $(dd\sigma)$=2, $\Delta$=6, and $U'$=20,
and investigate the dependence on $J$ and $V$.
The dependence on $(dd\sigma)$, $\Delta$, and $U'$
will be discussed elsewhere in future.

%%%%%%%%%%%%%%% Fig. 1 %%%%%%%%%%%%%%%%%%%%%%%%%%%%%%%%%%%%%
\begin{figure}[t]
\includegraphics[width=0.4\textwidth]{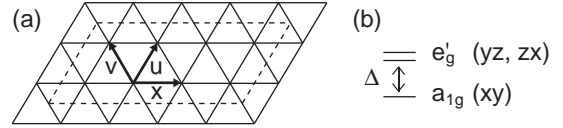}
\caption{%
(a) Triangular lattice with three directions depicted by arrows.
An eight-site cluster enclosed by dotted lines is considered
in our numerical calculations.
(b) Schematic view of level splitting among $t_{\rm 2g}$ orbitals.
}
\end{figure}
%%%%%%%%%%%%%%%%%%%%%%%%%%%%%%%%%%%%%%%%%%%%%%%%%%%%%%%%%%%%

Our result is summarized in Fig.~2(a), which is
the ground-state phase diagram
in the ($V,J$) plane.
Comparing the lowest energies in the subspaces of
the $z$ component of the total spin,
we determine the total spin in the ground state.
Then, we find that for a large set of parameters,
the ground state is a paramagnetic (PM) phase
characterized by total spin zero,
but there appear several types of PM phases
in terms of spin-charge-orbital configurations.
Among them, the phase I is characterized by a stripe,
in which ferromagnetic (FM) spin chains are coupled
with each other in an antiferromagnetic (AFM) manner,
as shown in the inset of phase I.
Note that the charge is uniformly distributed
in the background of ferro-orbital (FO) arrangement.

On the other hand, the phases II and III exhibit charge order.
To clarify what types of charge configurations emerge
in the present system,
we investigate the charge structure factor.
When we examine the $V$ dependence of the charge structure factor
for $J$=0, it is clearly observed that it develops abruptly
around at $V$=10,
indicating the stabilization of a specific charge structure.
The results suggest a charge stripe composed of one-dimensional
chains, as shown in the inset of phase II.
Namely, the spin frustration is removed due to charge ordering.
It should be noted that a similar stripe charge pattern
has been suggested for a CoO$_{2}$ triangular-lattice model
by Hartree-Fock calculations \cite{Mizokawa2004}.
Note that in the phase II, two of three orbitals are found
to be fully occupied due to the effect of FO arrangement,
similar to the phase I.

%$C({\bf q})=\sum_{{\bf i},{\bf j}}
%e^{i{\bf q}\cdot({\bf i}-{\bf j})}
%\langle \rho_{\bf i} \rho_{\bf j} \rangle/N$

%%%%%%%%%%%%%%% Fig. 2 %%%%%%%%%%%%%%%%%%%%%%%%%%%%%%%%%%%%%
\begin{figure}[t]
\includegraphics[width=0.46\textwidth]{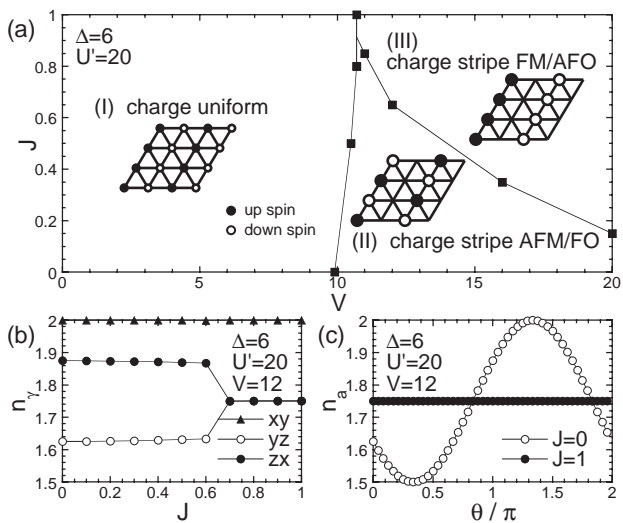}
\caption{%
(a) Ground-state phase diagram in the ($V,J$) plane for
$\Delta$=6 and $U'$=20 at electron density 5.5.
Inset denotes schematic view of the charge stripe and spin
configuration in each phase.
The size of circle indicates the degree of the charge density.
(b) Electron densities in the $xy$, $yz$, and $zx$ orbitals.
(c) Electron density in the $a$ orbital,
represented by the linear combination of the $yz$ and $zx$ orbitals.
}
\end{figure}
%%%%%%%%%%%%%%%%%%%%%%%%%%%%%%%%%%%%%%%%%%%%%%%%%%%%%%%%%%%%

Now we consider the effect of $J$ on the spin structure
in the stripe charge state.
For this purpose, it is useful to measure
the spin structure factor.
For $V$=12, we observe a transition in the spin structure
factor at around $J$=0.6,
but the stripe charge configuration is not changed.
For $J$$<$0.6, we find AFM correlations
in each one-dimensional charge stripe,
as shown in the inset of phase II.
On the other hand, for $J$$>$0.6,
the spin configuration changes to a FM one,
as shown in the inset of phase III.
Weak antiferro-orbital (AFO) correlations
exist in the chains, since the electrons can
gain kinetic energy due to the so-called double-exchange
mechanism.

%$S({\bf q})=\sum_{{\bf i},{\bf j}}
%e^{i{\bf q}\cdot({\bf i}-{\bf j})}
%\langle S_{\bf i}^{z} S_{\bf j}^{z} \rangle/N$
%with
%$S_{\bf i}^z=\sum_{\gamma}
%(\rho_{{\bf i}\gamma\uparrow}-\rho_{{\bf i}\gamma\downarrow})/2$

In order to identify the orbital arrangement,
it is convenient to evaluate 
the electron densities,
$n_{\gamma}=\sum_{\bf i}\langle \rho_{{\bf i}\gamma} \rangle/N$,
in the $xy$, $yz$, and $zx$ orbitals.
In Fig.~2(b), we show $n_{\gamma}$ vs. $J$ for $V$=12.
It is observed that the lower $xy$ orbital is fully occupied
irrespective of $J$,
while the electron densities in the $yz$ and $zx$ orbitals change
suddenly at around $J$=0.6.
For $J$$<$0.6, electrons favorably occupy the $zx$ orbital,
but the $zx$ orbital is not fully occupied.
On the other hand,
the $yz$ and $zx$ orbitals are equally occupied for $J$$>$0.6.

However, in order to determine the orbital structure,
we need to pay due attention to the definition of orbitals.
Here, by using an angle $\theta_{\bf i}$
to characterize the orbital shape,
we introduce operators for $a$ and $b$ orbitals such as
$d_{{\bf i}a\sigma}=
e^{i\theta_{\bf i}/2}
[\cos(\theta_{\bf i}/2)d_{{\bf i},yz,\sigma}+
\sin(\theta_{\bf i}/2)d_{{\bf i},zx,\sigma}]$
and
$d_{{\bf i}b\sigma}=
e^{i\theta_{\bf i}/2}
[-\sin(\theta_{\bf i}/2)d_{{\bf i},yz,\sigma}+
\cos(\theta_{\bf i}/2)d_{{\bf i},zx,\sigma}]$.
In Fig.~2(c),
we show the electron density in the $a$ orbital
with $\theta_{\bf i}$=$\theta$.
We find that for $J$=0, the $a$ orbital of $\theta$=$4\pi/3$
%(equivalently the $b$ orbital of $\theta=\pi/3$)
is fully occupied.
Namely, only one orbital becomes relevant for small $J$.
This FO arrangement is also found in the phase I.
On the other hand, for $J$=1,
the electron density does $not$ change 
even when $\theta$ is varied,
indicating that two orbitals remain active.

Note that we have suggested the stripe charge order
from the 8-site calculations, but the shape of the stripe may
change from straight to zigzag.
To confirm this point,
it is necessary to perform the calculations in larger-sized systems,
for instance, using a model including only $e'_{\rm g}$ orbitals.
This is among possible future developments.

In summary, in the PM phase,
one or two orbitals among the $t_{\rm 2g}$ orbitals become relevant
due to the effect of orbital arrangement.
The charge stripe stabilized by $V$ consists of
AFM/FO chains for small $J$,
while it changes to FM/AFO ones with increasing $J$.

We thank E. Dagotto for discussions.
T.H. is supported by the Japan Society for the Promotion of Science
and by the Ministry of Education, Culture, Sports, Science,
and Technology of Japan.

%%%%%%%%%%%%%%%%%%%%%%%%%%%%%%%%%%%%%%%%%%%%%%%%%%%%%%%%%%%%

\end{document}